\documentclass[letterpaper,11pt]{article}
\pdfoutput=1

\usepackage{jheppub}
\usepackage{mathtools}

\newcommand{\RN}[1]{
  \textup{\uppercase\expandafter{\romannumeral#1}}
}
\newcommand\numberthis{\addtocounter{equation}{1}\tag{\theequation}}

\def\d{\delta}
\def\e{\epsilon}
\def\D{\Delta}
\def\O{\mathcal{O}}

\newcommand{\zb}{\bar{z}}
\newcommand{\hypf}{{}_2F_{1}}

\title{Four-point correlators of light-ray operators in CCFT}
\author[a,b]{Yangrui Hu,}
\author[a]{Luke Lippstreu,}
\author[a,b]{Marcus Spradlin,}
\author[a]{Akshay Yelleshpur Srikant,}
\author[a]{Anastasia Volovich}

\affiliation[a]{Department of Physics,
	Brown University,
	Providence, RI 02912, USA}
\affiliation[b]{Brown Theoretical Physics Center,
	Box S, 340 Brook Street, Barus Hall,
	Providence, RI 02912, USA}

\emailAdd{yangrui\_hu@brown.edu}
\emailAdd{luke\_lippstreu@brown.edu}
\emailAdd{marcus\_spradlin@brown.edu}
\emailAdd{akshay\_yelleshpur\_srikant@brown.edu}
\emailAdd{anastasia\_volovich@brown.edu}

\abstract{
We compute the four-point correlator of two gluon light-ray operators and two gluon primaries from the four-gluon celestial amplitude
in $(2,2)$ signature spacetime.
The correlator is non-distributional and allows us to verify that
light-ray operators appear in the OPE of two gluon primaries.  We also carry out a conformal block decomposition of the terms involving the exchange of gluon operators.}

\begin{document}
	\maketitle
	\section{Introduction}
	Correlation functions in conformal field theories contain a wealth of information.  In generic CFTs, the OPE coefficients are related to three-point functions and four-point functions contain information about the spectrum of the theory, which can be deduced by means of the conformal block decomposition. Celestial conformal field theories (CCFTs), whose three- and (tree-level) four-point correlators are easily computed via Mellin\footnote{Massive CCFT correlators are computed by convolution with bulk-to-boundary propagators instead of Mellin transforms.} transforms of momentum space scattering amplitudes~\cite{Pasterski:2016qvg, Pasterski:2017kqt, Pasterski:2017ylz}, make these relationships opaque due to the distributional nature of their correlators. Nevertheless, several attempts have been made to deduce the spectrum and the OPE coefficients. These include direct analysis of the distributional correlators using conventional CFT techniques~\cite{Nandan:2019jas, Atanasov:2021cje, Guevara2021tvr, Fan:2022vbz}, asymptotic symmetries~\cite{Pate:2019lpp, Himwich:2021dau, Jiang:2021ovh}, and representation theory~\cite{Banerjee:2019aoy, Banerjee:2019tam, Pasterski:2020pdk, Pasterski:2021fjn}. The analysis has revealed that the CCFT spectrum typically contains light-ray operators and shadows in addition to the usual primaries.
	
	Therefore the computation of correlators involving shadow or light-ray operators is essential to gain a better understanding of the role played by these operators in CCFT. At four points, the additional integrals inherent in the definitions of the shadow or light transforms render the correlators
	non-distributional, thereby making the analysis more straightforward (or, at least, more ``traditional''). The shadow transform has been employed in~\cite{Fan:2021isc, Fan:2021pbp} to obtain, amongst other things, a non-distributional four-point correlator and its conformal block decomposition. The light transform is most naturally defined in a Lorentzian CFT,\footnote{For some discussion on the relevance of the light transform directly on the celestial sphere see~\cite{Donnay:2022sdg}.} which requires an analytic continuation of the Euclidean CFT on the celestial sphere. It has been shown that such an analytically continued CFT lives on a Lorentzian torus~\cite{Atanasov:2021oyu} and the correlators have a natural interpretation as Mellin transforms of scattering amplitudes in $(2,2)$ signature bulk spacetime. These techniques have been employed in~\cite{Sharma:2021gcz} to produce
	non-distributional three-point functions.

There are other motivations for studying light-ray operators in CFT beyond the desire to better understand their role in OPEs.  It was noted in~\cite{Sharma:2021gcz} that the light transform is an analogue of Witten's half-Fourier transform to twistor space~\cite{Witten:2003nn}, and the generalization where one transforms each operator on $z_i$ or $\bar{z}_i$ depending on helicity is similarly analogous to the ``link representation'' of~\cite{Arkani-Hamed:2009hub,Arkani-Hamed:2009ljj}; these developments have had enormous impact on the study of amplitudes. Moreover, in~\cite{Strominger:2021mtt} Strominger recognized a universal symmetry algebra based on $w_{1+\infty}$ in the gravitational S-matrix by carrying out an appropriate light transform on the result of~\cite{Guevara:2021abz}.

\smallskip

The central result of our paper is the following formula for the tree-level correlator of two gluon light-ray operators $\bar{\bf L}[\mathcal{O}{}_{\Delta,J}]$ and two gluon primaries $\mathcal{O}_{\Delta,J}$:
\begin{multline}
\langle \bar{\bf L}[\mathcal{O}{}_{\Delta_1,-}](z_1,\bar{z}_1)
\bar{\bf L}[\mathcal{O}{}_{\Delta_2,-}](z_2,\bar{z}_2)\,\mathcal{O}_{\Delta_3,+}(z_3,\bar{z}_3)\mathcal{O}_{\Delta_4,+}(z_4,\bar{z}_4)\rangle\\
=\pi\d(\beta)\,{\rm sgn}\left(\frac{z}{z-1}\right)
\frac{|z|^{\frac{4}{3}-\frac{\D_1+\D_2}{2}}|1-z|^{\frac{1}{3}-\frac{\D_1+\D_4}{2}}\mathcal{F}(z,\zb)}{|\bar{z}_{13}|^{\D_3-1}|\bar{z}_{14}|^{\D_2+\D_4-2}|\bar{z}_{24}|^{1-\D_2}}\prod_{i<j}^{4}|z_{ij}|^{\frac{2}{3}-\frac{\D_i+\D_j}{2}-\frac{J_i+J_j}{2}}
\label{eq:result}
\end{multline}
where
\begin{align}
\beta=\sum_{i=1}^4 \Delta_i-4\,,
\qquad z_{ij} = z_i-z_j\,,
\qquad z = \frac{z_{12} z_{34}}{z_{13} z_{24}}\,.
\label{eq:betazdef}
\end{align}
In the region $z, \zb>1$, the function $\mathcal{F}$ is given by
	\begin{multline}
	    \mathcal{F}\left(z,\zb\right)  =
	    C(\Delta_1,\Delta_2)
			|z-1|^{\Delta_4-2}|z-\bar{z}|^{\Delta_1+\Delta_2-1}\,_2F_1\left[2-\Delta_4,\Delta_1,\Delta_1+\Delta_2,\frac{z-\bar{z}}{z-1}\right]\\
			+
	    C(\Delta_3-1,\Delta_4-1)
				|z-1|^{1-\Delta_3}
				{}_2F_1\left[1-\Delta_2,\Delta_3-1,\Delta_3+\Delta_4-2,\frac{z-\bar{z}}{z-1}\right],
				\label{eq:FregionB}
	\end{multline}
where
\begin{align}
	C(a,b)=B(a,b)+B(a,1-a-b)+B(b,1-a-b)\,, \qquad B(a,b) = \frac{\Gamma(a) \Gamma(b)}{\Gamma(a+b)}\,.
	\label{eq:Cdef}
\end{align}
A complete description of $\mathcal{F}$ for all values of $z, \zb$ can be found in Section~\ref{sec:lighttransform}.
	
The paper is organized as follows. In Section~\ref{sec:4pt22} we discuss the tree-level four-gluon amplitude in $(2,2)$ signature spacetime and compute the corresponding celestial correlator. We highlight key differences compared to its $(3,1)$ analogue. In Section~\ref{sec:lighttransform} we compute its double light transform and derive the result~(\ref{eq:result}) for all values of $z,\zb$. We extract information about the OPE of gluon primaries from the correlator in Section~\ref{sec:collinearlimits}, and we study its conformal block decomposition in Section~\ref{sec:conf-block-decomp}.

	\section{\texorpdfstring{The four-point celestial amplitude in $(2,2)$ signature}{The four-point celestial amplitude in (2,2) signature}}
	\label{sec:4pt22}

 Any (non-zero) null four-vector in $(2,2)$ signature can be uniquely parameterized as
	\begin{align}
	p^{\mu} = \epsilon \omega \left( 1+ z \zb, z + \zb, z - \zb, 1- z \zb \right)
	\end{align}
	where $\epsilon=\pm 1$, $\omega>0$, and $z$ and $\zb$ are independent real variables.
	In $(3,1)$ signature $\bar{z}$ would be the complex conjugate of $z$ and $\epsilon$ would indicate whether $p^\mu$ describes an incoming or outgoing gluon. In $(2,2)$ signature we do not have this interpretation; rather $\epsilon$ labels different Poincar{\'e} patches. Note that it transforms covariantly under $SL(2,\mathbb{R}) \times SL(2,\mathbb{R})$ conformal transformations: $\e\rightarrow \e\, \textup{sgn}((cz+d)(\bar{c}\bar{z}+\bar{d}))$ when $z \to (a z+b)/(c z+d)$ and $\bar{z} \to (\bar{a} \bar{z} + \bar{b})/(\bar{c} \bar{z} + \bar{d})$.
	
	 The tree-level, color-ordered, four-gluon amplitude (with the helicities of gluons 1 and 2 being negative and those of 3 and 4 positive) is given by the Parke-Taylor formula\footnote{We suppress an overall factor proportional the square of the coupling constant.}
	\begin{equation}
			\mathcal{A}_{--++}(\omega_i,z_i,\bar{z}_i,\e_i) = \frac{z_{12}^3}{z_{23}z_{34}z_{41}} \frac{\omega_1 \omega_2}{\omega_3 \omega_4 } \, \delta^{4} \left(\sum_{i=1}^4 p^{\mu}_i \right).
			\label{eq:4ptamp}
	\end{equation}
	The corresponding celestial amplitude, obtained by Mellin transforming\footnote{The Mellin integral is initially defined for $\Delta_i = 1 + i \mathbb{R}$, and then understood to be defined for more general $\Delta_i$ by analytic continuation.} on $\omega_i$, is
\begin{align}
\label{eq:4ptcelestial}
       & \tilde{\mathcal{A}}_{--++}(\Delta_i,z_i,\bar{z}_i,\e_i) ~=~ \Big(\prod_{i=1}^4\,\int_0^{\infty}\,d\omega_i\,\omega_i^{\Delta_i-1}\Big)\, \mathcal{A}_{--++}(\omega_i,z_i,\bar{z}_i,\e_i)\nonumber\\
			&\quad =\frac{\pi}{2}
			\frac{ \d(\beta)\delta(z-\bar{z})|z_{12}|^3 }{ | z_{23}z_{34}z_{41} z_{13}z_{24} \bar{z}_{13} \bar{z}_{24} | }
			\left|\frac{z_{24}
\Bar{z}_{24}}{z_{12} \Bar{z}_{12}}z\right|^{\D_1}\left|\frac{ \Bar{z}_{3
4} z_{34}}{z_{23}\Bar{z}_{23}}\frac{1-z}{z}\right|^{\D_2}\left|\frac{\Bar{z}_{24} z_{24}}{ z_{23}\Bar{z}_{23}}(z-1)\right|^{\D_3-2}\\
			&\quad\ \times{\rm sgn}\left(\frac{z}{z-1}\right)\Theta\left(-\epsilon_1\epsilon_4\frac{z_{24}\Bar{z}_{24}}{z_{12} \Bar{z}_{12}}z\right)\Theta\left(\e_2\e_4\frac{ {z}_{34} \bar{z}_{34}}{z_{23}\Bar{z}_{23}}\frac{1-z}{z}\right)\Theta\left(\e_3\e_4\frac{{z}_{24} \Bar{z}_{24}}{ z_{23}\Bar{z}_{23}}(z-1)\right),\nonumber
\end{align}
where $\Theta(x)$ denotes the Heaviside step function. The $(3,1)$ signature analogue of this result was presented in (3.6) of~\cite{Pasterski:2017ylz}. It is easy to see that the indicator functions present in that equation can be rewritten as $\Theta$ functions depending solely on the cross-ratio. In contrast, the $\Theta$ functions in~(\ref{eq:4ptcelestial}) cannot be simplified further owing to the fact sgn$(z_{ij}\zb_{ij})$ is not fixed in $(2,2)$ signature spacetime. The $\Theta$ functions also complicate the analytic continuation of~(\ref{eq:4ptcelestial}) to arbitrary values of $z_i, \zb_i$. A simple way to circumvent this issue is to define\footnote{A similar approach of summing over channels was taken in~\cite{Fan:2021isc}, albeit in $(3,1)$ signature. See~\cite{Chang:2021wvv} for an alternative procedure for analytically continuing celestial amplitudes to the entire $z,\bar{z}$ plane.}
\begin{equation}
    \begin{split}
        & \langle \mathcal{O}{}_{\Delta_1,-}(z_1,\bar{z}_1)
			\mathcal{O}{}_{\Delta_2,-}(z_2,\bar{z}_2)\,\mathcal{O}_{\Delta_3,+}(z_3,\bar{z}_3)\mathcal{O}_{\Delta_4,+}(z_4,\bar{z}_4)\rangle ~\coloneqq~ \sum_{\e_i=\pm}\tilde{\mathcal{A}}_{--++}(\Delta_i,z_i,\bar{z}_i,\e_i)
	\end{split}
\end{equation}
where each operator $\O_{\D, J}$ can be thought of as $\O^{\rm{``in"}}_{\D,J} + \O^{\rm{``out"}}_{\D,J}$. Here $\D$ is conformal weight and $J$ is spin, which we take to be equal to the helicity of the particle (i.e., $\pm 1$ for gluons). The resulting expression is free of $\Theta$ functions due to the identity
\begin{equation}
  \sum_{\e_i=\pm 1}  \Theta\left(-\e_1\e_4\frac{z_{24}\Bar{z}_{24}}{z_{12} \Bar{z}_{12}}z\right)\Theta\left(\e_2\e_4\frac{ {z}_{34} \bar{z}_{34}}{z_{23}\Bar{z}_{23}}\frac{1-z}{z}\right)\Theta\left(\e_3\e_4\frac{{z}_{24} \Bar{z}_{24}}{ z_{23}\Bar{z}_{23}}(z-1)\right)=2\,,
\label{eq:killthetas}
\end{equation}
Note that the sum receives contributions from configurations with one, two, or three $\epsilon$'s being positive (and the others negative); this contrasts with intuition from $(3,1)$ signature where valid kinematic configurations exist only when precisely two $\epsilon$'s are positive.
Applying~(\ref{eq:killthetas}) and generously using the delta function $\delta(z-\bar{z})$, we find that the correlator can be put into the form
\begin{multline}
        \langle \mathcal{O}{}_{\Delta_1,-}(z_1,\bar{z}_1)
			\mathcal{O}{}_{\Delta_2,-}(z_2,\bar{z}_2)\,\mathcal{O}_{\Delta_3,+}(z_3,\bar{z}_3)\mathcal{O}_{\Delta_4,+}(z_4,\bar{z}_4)\rangle\\
		= \pi\d(\beta)\delta(z-\bar{z}){\rm sgn}\left(\frac{z}{z-1}\right)\frac{|z|^{\frac{5}{3}}}{{|1-z|^{\frac{1}{3}}}}\prod_{i<j}^{4}|z_{ij}|^{\frac{2}{3}-\frac{\D_i+\D_j}{2}-\frac{J_i+J_j}{2}}|\bar{z}_{ij}|^{\frac{2}{3}-\frac{\D_i+\D_j}{2}+\frac{J_i+J_j}{2}}\,.
    \label{eq:startingamplitude}
\end{multline}
Note that the terms in the product are required by conformal invariance, which does not fix the overall dependence on the cross-ratio $z$. We emphasize that the absolute values in~(\ref{eq:startingamplitude}) follow directly from our starting point~(\ref{eq:4ptamp}) in $(2,2)$ signature; they are not imposed by hand.
However, it is worth pointing out that the absolute values obscure all information about causality. Indeed the causal structure of correlation functions is encoded in branch cuts which arise as we cross the light-cone singularities at $z_{ij}=0$ or $\zb_{ij}=0$. We hope to analyze these issues in more detail in the future.

\section{The light transform}
\label{sec:lighttransform}
The ``anti-holomorphic'' light transform of an operator $\mathcal{O}_{\D,J}$ with conformal weight $\D$ and spin $J$ is defined as
\begin{equation}
\label{eq:lightrayoperatordef}
    {\bar{\bf{L}}}[\mathcal{O}_{\D,J}](z,\bar{z})\coloneqq \int_{-\infty}^{\infty}\frac{d \bar{z}'}{|\bar{z}'-\bar{z}|^{2-\D+J}}\mathcal{O}_{\D,J}(z,\bar{z}')\,.
\end{equation}
It is easy to check that $ {\bar{\bf{L}}}[\mathcal{O}_{\D,J}](z,\bar{z})$ transforms as an operator with conformal weight $1+J$ and spin $\D-1$\footnote{This definition is satisfactory for the purposes of this paper since we only consider operators without  ``incoming"  or ``outgoing" $\epsilon$ labels. A definition appropriate for in or out operators was given in~\cite{Sharma:talk2022}.}. A similar definition exists for the ``holomorphic'' light transform with respect to $z$, which we will denote by ${\bf{L}}[\mathcal{O}_{\D,J}](z,\bar{z})$. For more details, we refer the reader to~\cite{Kravchuk:2018htv, Gelfand:105396}.

We  now compute the light transforms of the correlator~(\ref{eq:startingamplitude}). The computation of the first light transform is straightforward due to the presence of the delta function, which we write as
\begin{equation}
	\delta\left(z-\Bar{z}'\right) ~=~ \frac{|\bar{z}_{23}\bar{z}_{24}\bar{z}_{34}|}{|\bar{z}_{34}-z\bar{z}_{24}|^2}\,\delta\left(\bar{z}_1'-\frac{\bar{z}_{34}\bar{z}_2-z\bar{z}_{3}\bar{z}_{24}}{\bar{z}_{34}-z\bar{z}_{24}}\right)
\end{equation}
to obtain
\begin{equation}
    \begin{split}
        &\langle {\bar{\bf{L}}}[\mathcal{O}{}_{\Delta_1,-}](z_1,\bar{z}_1)
			\mathcal{O}{}_{\Delta_2,-}(z_2,\bar{z}_2)\,\mathcal{O}_{\Delta_3,+}(z_3,\bar{z}_3)\mathcal{O}_{\Delta_4,+}(z_4,\bar{z}_4)\rangle\\
			&\qquad =\int\frac{\text{d}\bar{z}_{1}'}{|\bar{z}_{1'1}|^{1-\Delta_1}}\langle \mathcal{O}{}_{\Delta_1,-}(z_1,\bar{z}_{1}')
			\mathcal{O}{}_{\Delta_2,-}(z_2,\bar{z}_2)\,\mathcal{O}_{\Delta_3,+}(z_3,\bar{z}_3)\mathcal{O}_{\Delta_4,+}(z_4,\bar{z}_4)\rangle\\
			&\qquad=\pi\d(\beta)\,{\rm sgn}\left(\frac{z}{z-1}\right)\left(\prod_{i<j}^{4}|z_{ij}|^{\frac{2}{3}-\frac{\D_i+\D_j}{2}-\frac{J_i+J_j}{2}}\right)\,|z|^{\frac{4}{3}-\frac{\D_1+\D_2}{2}}\\
			&\qquad\qquad\times |\bar{z}_{13}|^{\D_1-1}|\bar{z}_{23}|^{\D_4-2}|\bar{z}_{24}|^{1-\D_2-\D_4}|\bar{z}_{34}|^{\D_2}|1-z|^{\frac{1}{3}-\frac{\D_1+\D_4}{2}}|z-\bar{z}|^{\D_1-1}\,.
    \end{split}
    \label{eq:firstlighttransform}
\end{equation}
It is worthwhile to pause here to draw attention to the bulk point singularity located at $z=\zb$. While such singularities have been shown to absent in correlation functions of {\it local} operators in~\cite{Maldacena:2015iua}, their presence in CCFT has already been hinted at in~\cite{Lam:2017ofc}.

We can proceed with the computation of the second light transform in a similar manner. We choose to light transform the remaining negative helicity gluon w.r.t $\zb_2$. Making use of~(\ref{eq:firstlighttransform}) we find
	\begin{equation}
		\begin{split}
			&\langle \bar{\bf L}[\mathcal{O}{}_{\Delta_1,-}](z_1,\bar{z}_1)
			\bar{\bf L}[\mathcal{O}{}_{\Delta_2,-}](z_2,\bar{z}_2)\,\mathcal{O}_{\Delta_3,+}(z_3,\bar{z}_3)\mathcal{O}_{\Delta_4,+}(z_4,\bar{z}_4)\rangle\\
			&\qquad~=~\int_{-\infty}^{\infty}\, \frac{d\bar{z}_2'}{|\bar{z}_{2'2}|^{1-\Delta_2}}\langle  \bar{\bf L}[\mathcal{O}{}_{\Delta_1,-}](z_1,\bar{z}_1)
			\mathcal{O}{}_{\Delta_2,-}(z_2,\bar{z}_{2}')\,\mathcal{O}_{\Delta_3,+}(z_3,\bar{z}_3)\mathcal{O}_{\Delta_4,+}(z_4,\bar{z}_4)\rangle\\
				&\qquad~=~\ \pi\d(\beta)\,{\rm sgn}\left(\frac{z}{z-1}\right)\Bigg(\prod_{i<j}^{4}|z_{ij}|^{\frac{2}{3}-\frac{\D_i+\D_j}{2}-\frac{J_i+J_j}{2}}\Bigg)\\
				&\hspace{1.7cm}|z|^{\frac{4}{3}-\frac{\D_1+\D_2}{2}}|1-z|^{\frac{1}{3}-\frac{\D_1+\D_4}{2}}\,|\bar{z}_{13}|^{1-\D_3}|\bar{z}_{14}|^{2-\D_2-\D_4}|\bar{z}_{24}|^{\D_2-1} \,\, 	{\cal F}(z,\bar{z})\,,
		\end{split}
		\label{eq:secondlighttransform-def}
	\end{equation}
where, with the help of the change of variable to
$t={\bar{z}_{12'}\bar{z}_{34}}/{\bar{z}_{13}\bar{z}_{2'4}}$, we
have
	\begin{equation}
		{\cal F}(z,\bar{z})~:=~ \int_{-\infty}^\infty dt\,|\bar{z}-t|^{\D_2-1}|z-t|^{\D_1-1}|1-t|^{\D_4-2}\label{eq:4markedpoints}
	\end{equation}
which is an integral over four marked points (one of which is at infinity).
We relegate the details of the evaluation of this integral to  Appendix~\ref{app:fourmarkedpointsintegral}. The result depends on the relative positions of $z,\zb$ and 1.  If they are on opposite sides of
1 ($z<1<\bar{z}$ or $\bar{z}<1<z)$ then the result can be written as
	 	\begin{multline}
	  {\cal F}(z,\bar{z}) =|1-\bar{z}|^{1-\Delta_3} C(\Delta_2,\Delta_3-1) \,_2F_1\left[1-\Delta_1,\Delta_3-1,\Delta_2+\Delta_3-1,\frac{1-z}{1-\bar{z}}\right]\\
    + \frac{|1-z|^{\Delta_1+\Delta_4-2}}{|1-\bar{z}|^{1-\Delta_2}}C(\Delta_1,\Delta_4-1)\,_2F_1\left[1-\Delta_2,\Delta_4-1,\Delta_1+\Delta_4-1,\frac{1-z}{1-\bar{z}}\right],\label{eq:FregionA}
	\end{multline}
	where $C$ is defined in~(\ref{eq:Cdef}).
	 On the other hand, if they are on the same side of 1 (either $z,\bar{z}>1$ or
	 $z,\bar{z} <1$) then the result takes the form shown in~(\ref{eq:FregionB}).

\section{Collinear limits and the OPE}
	\label{sec:collinearlimits}
Four-point correlators in any CFT contain information about the OPE of the operators they involve. A direct computation of the OPE from the four-gluon correlator in CCFT is usually hindered by the fact that the correlator is distributional, proportional to $\delta(z-\bar{z})$\footnote{OPE coefficients have been extracted from the four-point function prior to Mellin transformation in~\cite{Banerjee:2020kaa,Ebert:2020nqf}.}. In this section, we  exploit the non-distributional nature of the correlator involving two light-ray operators~(\ref{eq:secondlighttransform-def}) to obtain information about the OPE between gluon primaries.
To that end we consider the collinear limit as both
$z_{34}$ and $\bar{z}_{34}$ approach zero.
In this limit the cross-ratio $z$ approaches zero, and from
the appropriate expression for ${\cal F}(z,\bar{z})$ given in~(\ref{eq:FregionB}) we begin by reading off the leading
term as $z_{34} \to 0$:
			\begin{equation}
		\begin{split}
			&  \langle \bar{\bf L}[\mathcal{O}{}_{\Delta_1,-}](z_1,\bar{z}_1)
			\bar{\bf L}[\mathcal{O}{}_{\Delta_2,-}](z_2,\bar{z}_2)\,\mathcal{O}_{\Delta_3,+}(z_3,\bar{z}_3)\mathcal{O}_{\Delta_4,+}(z_4,\bar{z}_4)\rangle\\
			&\quad\sim \frac{\pi}{z_{34}}\,\d(\beta)\,{\rm sgn}(z_{12}z_{23}z_{31})
			|z_{12}|^{3-\D_1-\D_2}|z_{13}|^{\D_2-2}|z_{23}|^{\D_1-2}\\
			&\quad\quad\times\left(\frac{C(\D_3-1,\D_4-1)}{|\bar{z}_{13}|^{1-\D_1}|\bar{z}_{23}|^{1-\D_2}} \,_2F_1\left[1-\Delta_2,\Delta_3-1,\Delta_3+\Delta_4-2,\bar{z}\right]\right.\\
			&\quad\quad\quad+\left.\frac{|\bar{z}|^{\Delta_1+\Delta_2-1} C(\D_1,\D_2)}{|\bar{z}_{12}|^{1-\D_1-\D_2}|\bar{z}_{13}|^{\D_2}|\bar{z}_{23}|^{\D_1}|\bar{z}_{34}|^{\D_3+\D_4-3}} \,_2F_1\left[2-\Delta_4,\Delta_1,\Delta_1+\Delta_2,\bar{z}\right]
			\right),
		\end{split}
		\label{eq:34split}
	\end{equation}
where we have retained only the leading ${\mathcal{O}}(1/z_{34})$ singular terms.  It is now straightforward to take the limit $\bar{z}_{34}\to 0$.  The cross-ratio $\bar{z}$ becomes 0 in this limit, so the hypergeometric functions approach 1.  In terms of three-point functions with two or three light-ray operators, computed in~(\ref{equ:doubleL}) and~(\ref{equ:tripleL}) of Appendix~\ref{app:threeptlighttransforms}, the leading terms can be written as
\begin{multline}
        \langle \bar{\bf L}[\mathcal{O}{}_{\Delta_1,-}](z_1,\bar{z}_1)
			\bar{\bf L}[\mathcal{O}{}_{\Delta_2,-}](z_2,\bar{z}_2)\mathcal{O}_{\Delta_3,+}(z_3,\bar{z}_3)\mathcal{O}_{\Delta_4,+}(z_4,\bar{z}_4)\rangle\\
		\sim\frac{-1}{2 z_{34}} \left<\bar{\bf L}[\mathcal{O}_1{}^{}_{\Delta_1,-}]\bar{\bf L}[\mathcal{O}_{\Delta_2,-}]\left(C(\D_3-1,\D_4-1)\, \mathcal{O}_{\Delta_3+\D_4-1,+} +
			\frac{\bar{\bf L}[\mathcal{O}_{\Delta_3+\D_4-1,+}]}{|\bar{z}_{34}|^{\D_3+\D_4-3}}\right)\right>.
\end{multline}
By reinstating color indices and structure constants $f^{abc}$ of the gauge group in the obvious way,
we infer from this collinear limit the OPE
	\begin{align}
	   \mathcal{O}^a_{\Delta_i,+}(z_i,\bar{z}_i)\mathcal{O}^b_{\Delta_j,+}(z_j,\bar{z}_j)
\sim  \frac{-f^{abc}}{2 z_{ij}}\left(C(\D_i-1,\D_j-1)\,\mathcal{O}^c_{\D_i+\D_j-1}+
 \frac{\bar{\bf L}[\mathcal{O}^c_{\Delta_i+\D_j-1,+}]}{|\bar{z}_{ij}|^{\D_i+\D_j-3}}\right).
\label{equ:OPEansatz}
	\end{align}
The first term involves a gluon primary of weight $\D_i+\D_j-1$ and has been computed from various methods~\cite{Fan:2019emx, Pate:2019lpp, Jiang:2021csc, Adamo:2021lrv, Adamo:2021zpw, Himwich:2021dau}, while the second term involves a light-ray operator and was conjectured in Section~5 of~\cite{Guevara2021tvr}. The appearance of the second term is also consistent with the fact that the conformal block decomposition of four-point correlators involves the exchange of light-ray operators~\cite{Atanasov:2021cje}.

We pause here to point out that the OPE coefficient involving one ``incoming" and one ``outgoing" gluon computed in~\cite{Pate:2019lpp} by Mellin transforming the splitting function is proportional to  $B(\Delta_3-1,3-\Delta_3-\Delta_4)-B(\Delta_4-1,3-\Delta_3-\Delta_4)$. The relative minus sign between the two beta functions is apparently at odds with our result. However, the splitting function obtained from the $(2,2)$ signature amplitude~(\ref{eq:startingamplitude}) involves an absolute value and its Mellin transform is in agreement with~(\ref{equ:OPEansatz}).

The four-point correlator can also be used to compute the OPE between two light-ray operators and the OPE between one light-ray operator and a primary; we comment on these in Appendix~\ref{app:furtherlimits}.

	\section{Conformal block decomposition}\label{sec:conf-block-decomp}
	In the previous section we showed directly from the four-point correlator~(\ref{eq:secondlighttransform-def}) that the OPE of two primaries involves a linear combination of a primary and a light-ray operator. In this section we perform a conformal block decomposition of the term in~(\ref{eq:secondlighttransform-def}) corresponding to the exchange of gluon operators (meaning gluon primaries and their descendants, as opposed to light-ray operators and their descendants).
	
	The $SL(2,\mathbb{R})\times SL(2,\mathbb{R})$ conformal symmetry allows us to set $z_1=\infty, z_2 = 1, z_4=0$ (and similarly for $\zb_i$). Then
	$z = z_3$ and $\bar{z}=\bar{z}_3$, and we can extract
	\begin{multline}
		\lim_{z_1,\bar{z}_1\to \infty}|z_1|^{\Delta_1-1}|\bar{z}_1|^{1-\Delta_1}\langle \bar{\bf L}[\mathcal{O}_{\Delta_1,-}](z_1,\bar{z}_1)
			\bar{\bf L}[\mathcal{O}_{\Delta_2,-}](1,1)\,\mathcal{O}_{\Delta_3,+}(z,\bar{z})\mathcal{O}_{\Delta_4,+}(0,0)\rangle\\
			= -\pi\d(\beta)\,\Big[ {\cal I}_1(z,\bar{z})+{\cal I}_2(z,\bar{z})\Big],
		\end{multline}
	where  (we assume that $0<z,\bar{z}<1$)
	\begin{equation}
		\begin{split}
			{\cal I}_1(z,\bar{z}) ~=&~ \frac{C(\D_3-1,\D_4-1)}{z(1-z)^{\Delta_3}}\, _2F_1\left[1-\Delta_2,\Delta_3-1,\Delta_3+\Delta_4-2,\frac{\bar{z}-z}{1-z}\right],\\
			{\cal I}_2(z,\bar{z})~=&~\frac{C(\D_1,\D_2)}{z(1-z)^{3-\Delta_4}}\,|\bar{z}-z|^{\Delta_1+\Delta_2-1}\,_2F_1\left[2-\Delta_4,\Delta_1,\Delta_1+\Delta_2,\frac{\bar{z}-z}{1-z}\right].
		\end{split}
	\end{equation}
We focus on the first term, which corresponds to the exchange of gluon operators, and leave the decomposition of the second term, which corresponds to the exchange of light-ray operators, to future work.

We proceed along the lines of~\cite{Fan:2021isc} in order to massage ${\cal I}_1(z,\bar{z})$ into a form from which the conformal block decomposition can be read off.
	First  we use the identity
	\begin{equation}
		\begin{split}
	 \hypf\left(a,b_1,b_1+b_2,\frac{x-y}{1-y}\right) =	(1-y)^{a}F_1[a,b_1,b_2,b_1+b_2,x,y]\,,
		\end{split}
		\label{equ:AppellF1-simplify}
	\end{equation}
	and then  the Burchnall-Chaundy expansion~\cite{Identity1,identity2} of the Appell function
	\begin{multline}
			F_1[a,b_1,b_2,b_1+b_2,x,y] \\ =\sum_{n=0}^{\infty}\,\frac{(a)_n(b_1)_n(b_2)_n(c-a)_n}{n!(c+n-1)_n(c)_{2n}}\,x^n\,y^n\,_2F_1(a+n,b_1+n,c+2n,x) _2F_1(a+n,b_2+n,c+2n,y)\,,
		\label{equ:Appell-decomp}
	\end{multline}
	along with the Gauss recursion relations for $_2F_1$, to express
	\begin{equation}
		\begin{split}
			{\cal I}_1(z,\bar{z}) ~=&~ \sum_{n=0}^{\infty}\sum_{k=n}^{\infty}\,a_{k,n}\, K^{21}_{34}\left[\frac{\D_3+\D_4}{2}+k,\frac{\D_3+\D_4}{2}+n-1\right].\label{eq:finalconfblock}
		\end{split}
	\end{equation}
	Here the coefficients are
		\begin{multline}
			a_{k,n} = C(\D_3-1,\D_4-1)\,
			\frac{(1-\D_1)_n(1-\D_2)_n(\D_3-1)_n(\D_4-1)_n}{n!\,(\D_3+\D_4+n-3)_n(\D_3+\D_4-2)_{2n}}\\
			\times\sum_{m=0}^{k-n}\,\frac{(1-\D_2+n)_{m}(\D_4-1+n)_m}{(\D_3+\D_4+2n-2)_{2m}}\,\frac{(2-\D_1+n+m)_{k-n-m}(\D_3+n+m)_{k-n-m}}{(\D_3+\D_4-1+2n+2m)_{2k-2n-2m}}
			\label{eq:akncoef}
	\end{multline}
	and $K^{21}_{34}[h,\bar{h}]$ are the usual conformal blocks~\cite{Osborn:2012vt}
		\begin{equation}
	    K^{21}_{34}[h,\bar{h}]=z^{h-h_3-h_4}{}_2F_1[h-h_{12},h+h_{34},2 h,z]\bar{z}^{\bar{h}-\bar{h}_3-\bar{h}_4}{}_2F_1[\bar{h}-\bar{h}_{12},\bar{h}+\bar{h}_{34},2\bar{h},\bar{z}]\,,
	    \label{eq:blockdef}
	\end{equation}
	where $h_{ij} = h_i - h_j$,	$\bar{h}_{ij}=\bar{h}-\bar{h}_j$.
	In the present application, operators 3 and 4 are gluon primaries with $J=+1$ and for these we should take
	 $(h_i,\bar{h}_i) = (\frac{\Delta_i+J_i}{2}, \frac{\Delta-J}{2})$ in~(\ref{eq:blockdef}).  On the other hand, for the light-transformed operators 1 and 2 we need to take $h_{12} = \frac{\Delta_1-\Delta_2}{2} = - \bar{h}_{12}$ in terms of the original weights $\D_1, \D_2$ (the latter are the ones that appear in~(\ref{eq:akncoef})).

	We can read off the spectrum of exchanged states in (\ref{eq:finalconfblock}) to be
	\begin{equation}
	    (h,\bar{h})=\Big(\frac{\D_3+\D_4}{2}+k,\frac{\D_3+\D_4}{2}+n-1\Big),\quad k,n\in \mathbb{N}^{+},\,\, k\ge n
	\end{equation}
	which interestingly indicates that only positive helicity operators are exchanged.

	\acknowledgments
	We are grateful to Jorge Mago, Lecheng Ren, and Atul Sharma for useful
	comments and discussion.
	This work was supported in part by the US Department of Energy under contract {DE}-{SC}0010010 Task A and by Simons Investigator Award \#376208.
	The research of Y.~Hu is supported in part by the endowment from the Ford Foundation Professorship of Physics and the Physics Dissertation Fellowship provided by the Department of Physics at Brown University. Y.~Hu also acknowledges the support of the Brown Theoretical Physics Center.

	\appendix

	\section{The four marked point integral}\label{app:fourmarkedpointsintegral}
	In this Appendix we elaborate on how to evaluate the integral (\ref{eq:4markedpoints}),
	\begin{equation}
		{\cal F}(z,\bar{z}):= \int_{-\infty}^\infty dt\,|\bar{z}-t|^{\Delta_2-1}|z-t|^{\Delta_1-1}|1-t|^{\Delta_4-2}\,.\label{eq:mainintegralforF}
	\end{equation}
Besides the singular point at $|t|=\infty$, the integrand exhibits three singular points at $t=1,z,\bar{z}$. Thus there are seemingly six different configurations which need to be analyzed:
	\begin{alignat*}{2}
	 \RN{1}&: \bar{z}<1<z,\quad \RN{2}&&: z<1<\bar{z}\\
	    \RN{3}&: 1<z<\bar{z},\quad \RN{4}&&: 1<\bar{z}<z\\
	    \RN{5}&: \bar{z}<z<1,\quad \RN{6}&&: z<\bar{z}<1
	\end{alignat*}
We will demonstrate that the result of the integral~(\ref{eq:mainintegralforF}) can be brought to a form where there are only two distinct configurations. To see this, let us first evaluate $\mathcal{F}$ in configuration $\RN{1}:\bar{z}<1<z$. The integral then breaks up into four regions
	\begin{alignat*}{2}
			&{\cal F}(z,\bar{z}){\Big\vert}_{\RN{1}}\\
			&=\int_{-\infty}^{\bar{z}}dt(\bar{z}-t)^{\Delta_2-1}(z-t)^{\Delta_1-1}(1-t)^{\Delta_4-2}&&+\int_{\bar{z}}^{1}dt(t-\bar{z})^{\Delta_2-1}(z-t)^{\Delta_1-1}(1-t)^{\Delta_4-2}\\
			&+\int_{1}^{z}dt(t-\bar{z})^{\Delta_2-1}(z-t)^{\Delta_1-1}(t-1)^{\Delta_4-2} &&+ \int_{z}^{\infty}dt(t-\bar{z})^{\Delta_2-1}(t-z)^{\Delta_1-1}(t-1)^{\Delta_4-2}\,.\numberthis
		\label{equ:1<z<zbar}
	\end{alignat*}
	All of these four integrals are Gauss hypergeometric functions; explicitly
	\begin{align*}
	   &{\cal F}(z,\bar{z}){\Big\vert}_{\RN{1}} =|1-\bar{z}|^{1-\Delta_3}\,B(\Delta_2,\Delta_3-1) _2F_1\left[1-\Delta_1,\Delta_3-1,\Delta_2+\Delta_3-1,-\frac{|z-1|}{|1-\bar{z}|}\right]\\
	   &+|1-\bar{z}|^{\Delta_2+\Delta_4-2}|z-1|^{\Delta_1-1}\,B(\Delta_2,\Delta_4-1)\,_2F_1\left[1-\Delta_1,\Delta_4-1,\Delta_2+\Delta_4-1,-\frac{|1-\bar{z}|}{|z-1|}\right]\\
	   &+|z-1|^{\Delta_1+\Delta_4-2}|1-\bar{z}|^{\Delta_2-1}\,B(\Delta_1,\Delta_4-1)\,_2F_1\left[1-\Delta_2,\Delta_4-1,\Delta_1+\Delta_4-1,-\frac{|z-1|}{|1-\bar{z}|}\right]\\
	   &+|z-1|^{1-\Delta_3}\,B(\Delta_3-1,\Delta_1)\,_2F_1\left[1-\Delta_2,\Delta_3-1,\Delta_1+\Delta_3-1,-\frac{|1-\bar{z}|}{|z-1|}\right]\numberthis\label{eq:zb<1<z 4}
	\end{align*}
where we have maintained the ordering between the four regions in~(\ref{equ:1<z<zbar}) and the four terms in~(\ref{eq:zb<1<z 4}) to indicate to the reader which term arises from which region. Using standard identities and Euler transformations we can write the result more compactly
as~(\ref{eq:FregionA}).

 An explicit computation of the integral in configuration $\RN{2}$ and subsequent use of the abovementioned identities reveals the identical result. Similar techniques can be applied to show that in regions $\RN{3}-\RN{6}$ the integral takes the form~(\ref{eq:FregionB}).

\section{Three-point functions and their light transforms}	 \label{app:threeptlighttransforms}
	 In this Appendix we present expressions for three-point functions involving two and three light-ray operators. These expressions serve as a point of comparison for the OPE relations~(\ref{equ:OPEansatz}).
	
	 We begin by adapting the three gluon amplitude in $(2,2)$ signature spacetime presented in (3.5) of~\cite{Pasterski:2017ylz} to the context of our paper by performing a sum over the $\epsilon_i$. This results in
	 \begin{multline}
	          \langle \O_{\D_1,-}\left(z_1,\zb_1\right)\O_{\D_2,-}\left(z_2,\zb_2\right)\O_{\D_3,+}\left(z_3,\zb_3\right)\rangle\coloneqq \sum_{\e_1,\e_2,\e_3=\pm}	    \tilde{\mathcal{A}}_{--+}(\D_i,z_i,\bar{z}_i)\\
	    =-2\pi\delta\Big(3-\sum_{i=1}^3\D_i\Big)\frac{\textup{sgn}(z_{12}z_{23}z_{31})\delta(\bar{z}_{13})\delta(\bar{z}_{12})}{|z_{12}|^{-\D_3}|z_{23}|^{2-\D_1}|z_{13}|^{2-\D_2}}\,.
	     \end{multline}
A direct application of the definition of the light transform in~(\ref{eq:lightrayoperatordef}) yields
	\begin{multline}
			 \langle \bar{\bf L}[\mathcal{O}_{\Delta_1,-}]\left(z_1,\zb_1\right)\bar{\bf L}[\mathcal{O}_{\Delta_2,-}]\left(z_2,\zb_2\right)\,\mathcal{O}_{\Delta_3,+}\left(z_3,\zb_3\right)\rangle \\
			= -2\pi\d\Big(\sum_i\D_i-3\Big)\,{\rm sgn}(z_{12}z_{23}z_{31})\,\frac{|z_{12}|^{3-\D_1-\D_2}|z_{13}|^{\D_2-2}|z_{23}|^{\D_1-2}}{|\bar{z}_{13}|^{1-\D_1}|\bar{z}_{23}|^{1-\D_2}}
	\label{equ:doubleL}
	\end{multline}
	and
	\begin{multline}
			 \langle \bar{\bf L}[\mathcal{O}_{\Delta_1,-}]\left(z_1,\zb_1\right)\bar{\bf L}[\mathcal{O}_{\Delta_2,-}]\left(z_2,\zb_2\right)\bar{\bf L}[\mathcal{O}_{\Delta_3,+}]\left(z_3,\zb_3\right)\rangle \\
			= -2\pi\d\Big(\sum_i\D_i-3\Big)\,{\rm sgn}(z_{12}z_{23}z_{31})\,\frac{|z_{12}|^{3-\D_1-\D_2}|z_{13}|^{\D_2-2}|z_{23}|^{\D_1-2}}{|\bar{z}_{12}|^{1-\D_1-\D_2}|\bar{z}_{13}|^{\D_2}|\bar{z}_{23}|^{\D_1}}\,C(\D_1,\D_2)\,,
		\label{equ:tripleL}
	\end{multline}
	where $C$ is defined in~(\ref{eq:Cdef}).

\section{Further OPE limits of the four-point amplitude}
\label{app:furtherlimits}
	 In this Appendix we examine the $z_{12}\to 0$ and $z_{13} \to 0$ collinear limits of the correlator~(\ref{eq:secondlighttransform-def}). First consider the limit $z_{12} \to 0$. In this regime, the correlator becomes
	\begin{equation}
		\begin{split}
			&\langle \bar{\bf L}[\mathcal{O}{}_{\Delta_1,-}](z_1,\bar{z}_1)
			\bar{\bf L}[\mathcal{O}{}_{\Delta_2,-}](z_2,\bar{z}_2)\mathcal{O}_{\Delta_3,+}(z_3,\bar{z}_3)\mathcal{O}_{\Delta_4,+}(z_4,\bar{z}_4)\rangle\\
			&\qquad\qquad\sim  \frac{2\pi\,\d\left(\beta\right)\,{\rm sgn}(z_{12}z_{23}z_{34}z_{42})}{|z_{12}|^{\D_1+\D_2-3}}\,
			\Big(\frac{|\bar{z}_{23}|^{1-\D_3}|\bar{z}_{24}|^{1-\D_4}}{|z_{23}|^{\D_3}|z_{24}|^{\D_4}|z_{34}|}\,C(\D_3-1,\D_4-1)\\
			&\qquad\qquad\qquad\qquad~+~|\bar{z}_{12}|^{\D_1+\D_2-1}\,\frac{|\bar{z}_{23}|^{\D_4-2}|\bar{z}_{24}|^{\D_3-2}|\bar{z}_{34}|^{\D_1+\D_2-1}}{|z_{23}|^{\D_3}|z_{24}|^{\D_4}|z_{34}|}\,C(\D_1,\D_2)
			\Big).
		\end{split}\label{eq:l1l2OPE}
	\end{equation}
	This can be expressed in terms of generic three-point correlators by noting that
	the $z_{ij}$ and $\zb_{ij}$ dependence of each term reveals which types of operators must appear.  Schematically, we must have
	\begin{equation}
		\begin{split}
			&\langle \bar{\bf L}[\mathcal{O}{}_{\Delta_1,-}](z_1,\bar{z}_1)
			\bar{\bf L}[\mathcal{O}{}_{\Delta_2,-}](z_2,\bar{z}_2)\,\mathcal{O}_{\Delta_3,+}(z_3,\bar{z}_3)\mathcal{O}_{\Delta_4,+}(z_4,\bar{z}_4)\rangle\\
		&\sim ~\rho_1~\frac{\langle {\bf S}[\mathcal{O}^m_{\Delta_1+\D_2-1,-}](z_2,\bar{z}_2)\,\mathcal{O}_{\Delta_3,+}(z_3,\bar{z}_3)\mathcal{O}_{\Delta_4,+}(z_4,\bar{z}_4)\rangle}{|z_{12}|^{\D_1+\D_2-3}}\\
			&~+~\rho_2~\frac{|\bar{z}_{12}|^{\D_1+\D_2-1}}{|z_{12}|^{\D_1+\D_2-3}}\,C(\D_1,\D_2)\,\langle {\bf L}[\mathcal{O}^m_{\Delta_1+\D_2-1,-}](z_2,\bar{z}_2)\,\mathcal{O}_{\Delta_3,+}(z_3,\bar{z}_3)\mathcal{O}_{\Delta_4,+}(z_4,\bar{z}_4)\rangle\,,
		\end{split}
	\end{equation}
	where $\rho_1$ and $\rho_2$ are functions independent of $z,\zb$, and ${\bf S}[\mathcal{O}^m_{\Delta_1+\D_2-1,-}]$ and ${\bf L}[\mathcal{O}^m_{\Delta_1+\D_2-1,-}]$ are the shadow and light transforms of some massive operator. Since three-point correlators of massless operators are always distributional, by ``massive operator'' we simply mean one whose three-point functions are non-distributional. The appearance of such operators is necessary to account for the structures in~(\ref{eq:l1l2OPE}). It would be interesting to understand the physical content of these terms and make connections to the existing literature on light-ray OPEs (see for example~\cite{Kologlu:2019mfz}).
	
Finally, we consider the limit $z_{23}\to 0$, which corresponds to $z \to 1$. First it is helpful to use hypergeometric identities to rewrite the function $\mathcal{F}\left(z,\zb\right)$ as
	\begin{multline}
			{\cal F} \left(z, \zb\right) = \frac{C(\D_1,\D_4-1)}{|1-z|^{2-\D_1-\D_4}|z-\bar{z}|^{1-\D_2}}\,_2F_1\left[\,1-\D_2,\D_1,\D_1+\D_4-1,\frac{1-z}{\bar{z}-z}\,\right]\\
			+ \frac{C(\D_2,\D_3-1)}{|z-\bar{z}|^{1-\D_1}|1-\bar{z}|^{2-\D_2-\D_4}}\,_2F_1\left[\,1-\D_1,\D_2,\D_2+\D_3-1,\frac{1-z}{\bar{z}-z}\,\right]\,,
		\end{multline}
	which simplifies the limit $z \to 1$, giving
	\begin{equation}
		\begin{split}
			&\langle \bar{\bf L}[\mathcal{O}{}_{\Delta_1,-}](z_1,\bar{z}_1)
			\bar{\bf L}[\mathcal{O}{}_{\Delta_2,-}](z_2,\bar{z}_2)\,\mathcal{O}_{\Delta_3,+}(z_3,\bar{z}_3)\mathcal{O}_{\Delta_4,+}(z_4,\bar{z}_4)\rangle\\
			&\qquad\sim   \frac{\pi\,\d\left(\beta\right)\,{\rm sgn}(z_{12}z_{24}z_{41})}{z_{23}}\,\Big(C(\D_1,\D_4-1)\,\frac{|\bar{z}_{23}|^{\D_2-1}|\bar{z}_{12}|^{2-\D_2-\D_3}|\bar{z}_{14}|^{\D_1+\D_2+\D_3-3}}{|z_{12}|^{\D_1-2}|z_{24}|^{\D_4}|z_{14}| \,|z_{23}|^{\D_2+\D_3-1}}\\
			&\qquad+ C(\D_2,\D_3-1)\frac{|\bar{z}_{23}|^{1-\D_3} |\bar{z}_{14}|^{\D_1-1}|\bar{z}_{24}|^{\D_2+\D_3-2}}{|z_{12}|^{-\D_4}|z_{24}|^{\D_2+\D_3+\D_4-2}|z_{14}|^{3-\D_2-\D_3}|z_{23}|}\,\Big).
		\end{split}
	\end{equation}
	This can be rewritten in terms of generic three-point correlators as
	\begin{equation}
		\begin{split}
			& \langle \bar{\bf L}[\mathcal{O}^{}_{\Delta_1,-}](z_1,\bar{z}_1)
			\bar{\bf L}[\mathcal{O}^{}_{\Delta_2,-}](z_2,\bar{z}_2)\,\mathcal{O}_{\Delta_3,+}(z_3,\bar{z}_3)\mathcal{O}_{\Delta_4,+}(z_4,\bar{z}_4)\rangle\\
			~\sim&~ \rho'_1\,\frac{1}{z_{23}}\,\frac{|\bar{z}_{23}|^{\D_2-1}}{|z_{23}|^{\D_2+\D_3-1}}\,\langle \bar{\bf L}[\mathcal{O}{}_{\Delta_1,-}](z_1,\bar{z}_1)
			{\bf L}[\mathcal{O}^m_{\Delta_2+\D_3-1,+}](z_2,\bar{z}_2)\,\mathcal{O}_{\Delta_4,+}(z_4,\bar{z}_4)\rangle\\
			&~+~ \rho'_2\,\frac{C(\D_2,\D_3-1)}{z_{23}}\,\frac{|\bar{z}_{23}|^{1-\D_3}}{|z_{23}|}\,\langle \bar{\bf L}[\mathcal{O}{}_{\Delta_1,-}](z_1,\bar{z}_1)
			\bar{\bf L}[\mathcal{O}{}_{\Delta_2+\D_3-1,-}](z_2,\bar{z}_2)\,\mathcal{O}_{\Delta_4,+}(z_4,\bar{z}_4)\rangle\,,\label{23}
		\end{split}
	\end{equation}
	with two constants $\rho'_1$ and $\rho'_2$. The introduction of the massive light-ray operator is again necessary to produce a three-point correlator with the correct structure.

\bibliography{main3}
\bibliographystyle{JHEP}

\end{document}